\documentclass[12pt,preprint]{aastex} 

\usepackage{graphicx}

\newcommand{\halpha}{H$\alpha$}
\newcommand{\hbeta}{H$\beta$}
\newcommand{\hgamma}{H$\gamma$}

\newcommand{\usp}{\underbar{\hskip 7pt}}
\newcommand\av[1]{\langle#1\rangle}

\shortauthors{VANLANDINGHAM ET AL.}
\shorttitle{SDSS MAGNETIC WHITE DWARFS II.}

\received{}
\revised{}

\begin{document}

\tolerance 10000

\title{Magnetic White Dwarfs from the SDSS II. The Second and Third Data Releases\altaffilmark{1}}

\altaffiltext{1}{A portion of the results presented here were obtained with
the MMT Observatory, a facility operated jointly by The University of Arizona
and the Smithsonian Institution.}

\author{
Karen M. Vanlandingham\altaffilmark{2,3},
Gary D. Schmidt\altaffilmark{2},
Daniel J. Eisenstein\altaffilmark{2},
Hugh C. Harris\altaffilmark{4},
Scott F. Anderson\altaffilmark{5},
Pat B. Hall\altaffilmark{6},
James Liebert\altaffilmark{2},
Donald P. Schneider\altaffilmark{7},
Nicole M. Silvestri\altaffilmark{5},
Gregory S. Stinson\altaffilmark{5},
and
Michael A. Wolfe\altaffilmark{5}
}

\altaffiltext{2}{Steward Observatory, The University of Arizona, Tucson AZ
85721.} \email{kvanland@as.arizona.edu}
\altaffiltext{3}{Department of Astronomy, Columbia University, 550 West 120th Street, New York, NY 10027.}
\altaffiltext{4}{U.S. Naval Observatory, P.O. Box 1149, Flagstaff, AZ
86002-1149.}
\altaffiltext{5}{Department of Astronomy, University of Washington, Box 351580,
Seattle, WA 98195-1580}
\altaffiltext{6}{Department of Physics \& Astronomy, 128 Petrie Science \&
Engineering Building, York University, 4700 Keele St., Toronto, ON, M3J 1P3,
Canada}
\altaffiltext{7}{Pennsylvania State University, Department of Physics \&
Astronomy, 525 Davey Lab., University Park, PA 16802.}

\begin{abstract}

Fifty-two magnetic white dwarfs have been identified in spectroscopic
observations from the Sloan Digital Sky Survey (SDSS) obtained between mid-2002
and the end of 2004, including Data Releases 2 and 3.  Though not as numerous
nor as diverse as the discoveries from the first Data Release, the collection
exhibits polar field strengths ranging from 1.5~MG to $\sim$1000~MG, and
includes two new unusual atomic DQA examples, a molecular DQ, and five stars
that show hydrogen in fields above 500~MG.  The highest-field example,
SDSS~J2346+3853, may be the most strongly magnetic white dwarf yet discovered.
Analysis of the photometric data indicates that the magnetic sample spans the
same temperature range as for nonmagnetic white dwarfs from the SDSS, and
support is found for previous claims that magnetic white dwarfs tend to have
larger masses than their nonmagnetic counterparts. A glaring exception to this
trend is the apparently low-gravity object SDSS~J0933+1022, which may have a
history involving a close binary companion.

\end{abstract}


\section{Introduction}

In October 2004, the release of the third installment from the Sloan Digital
Sky Survey (SDSS DR3) brought the database current through June 2003, with
images to $g>22$ for nearly 5300 square degrees of the sky and fiber
spectroscopy for selected targets over almost 4200 square degrees (Abazajian et
al. 2004; 2005).  Though designed principally to probe the extragalactic
universe, the survey is providing a wealth of information on stellar
populations both on and off the main sequence, as well as an effective
discovery tool for stellar objects such as accretion binaries, brown dwarfs,
and other classes of unusual stars.  In this paper we focus on the magnetic
white dwarfs in the SDSS database.

The study of magnetic white dwarfs is important not only because it
is possible to detect magnetism in this stage over nearly 6 orders of
magnitude in strength, but it is generally thought that the fields on
white dwarfs are amplified versions of what pervaded the
main sequence progenitors.  Magnetic white dwarfs may therefore
prove to be critical in understanding the role(s) of magnetism in
earlier phases of stellar evolution, and provide a clue to the reason
behind the essentially universal presence of magnetic fields on neutron
stars.  Finally, magnetic white dwarfs provide our only means of
empirically studying the effects of fields $B\gtrsim10$~MG on atomic
and molecular emission processes.  Indeed, in virtually all cases,
observational discovery of magnetically shifted features on white
dwarfs has preceded the calculations that explain the splitting.
Wickramasinghe \& Ferrario (2000) provide an excellent review of the
state of field in the pre-SDSS era and amplify on the relevance to
broader questions of astrophysics.

The first paper in a series reporting the identification of isolated
(non-accreting) magnetic white dwarfs from the SDSS (Schmidt et al. 2003,
hereafter Paper I) presented 53 new stars through the first data release (DR1),
with magnetic fields in the range $1.5<B_p<560$~MG (MG = 10$^6$~G).  That list
nearly doubled the previously known sample of magnetic white dwarfs and
included 3 new magnetic DB (helium feature) stars as well as several with
exotic/unknown atmospheric compositions.  The current paper reports an
additional 52 discoveries from survey data available through the end of 2004,
including both DR2 and DR3.

\section{Observational Data}

The SDSS is compiling an enormous imaging and spectroscopic data set for
selected regions of the northern sky using a special-purpose 2.5 m telescope at
Apache Pt., New Mexico (e.g., Fukugita et al. 1996; Gunn et al. 1998; Lupton et
al. 1999; York et al. 2000; Lupton et al. 2001; Pier et al. 2003).  Magnetic
white dwarfs are recognized in the spectroscopic database, which targets
objects selected from deep, 5-color ($u,g,r,i,z$) images (Hogg et al. 2001;
Smith  et al. 2002; Ivezic et al. 2004).  Spectroscopic target selection is
optimized for QSOs and galaxies, and utilizes complex color criteria that have
evolved somewhat over time. This and the highly unusual spectra of strongly
magnetic white dwarfs imply that candidate spectra might be found in several
target classes. Indeed, based on their photometric colors, the objects reported
here were generally selected as QSO ($\sim$40\%) or HOT{\usp}STD
($\sim$20\%), with the remainder sprinkled among STAR{\usp}WHITE{\usp}DWARF,
STAR{\usp}BHB, and other categories lying off the stellar locus such as
SERENDIPITY{\usp}BLUE and SERENDIPITY{\usp}DISTANT (see Stoughton et al. 2002
for target category descriptions).  The spectroscopy is performed with twin
dual-beam spectrographs covering the regions $3900-6200$~\AA\ and
$5800-9200$~\AA\ and providing a resolving power ($\lambda/d\lambda\sim1800$)
sufficient to detect a Zeeman triplet at \halpha\ for $B\gtrsim1.5$~MG.

Additional optical spectroscopy and spectropolarimetry were obtained during the
period Feb.$-$Dec. 2004 for several unusual and questionable targets.   These
observations utilized the instrument SPOL (Schmidt et al. 1992) attached to the
Steward Observatory 2.3 m Bok telescope on Kitt Peak and the 6.5 m MMT atop Mt.
Hopkins.  In the configuration used, they provide a spectral coverage of
$\sim$$\lambda\lambda4200-8400$ and resolution $\sim$15~\AA.  Details of the
observing rationale can be found in Paper I.

\section{Spectroscopic Identification}

For Paper I, the recognition of magnetic objects among the many thousands of
survey spectra was carried out by eye, searching among a variety of target
categories of stars and unusual objects.  Additional candidate magnetic
objects were contributed by workers in other areas of SDSS research.  With
subjective inspection playing such a central role, questions of completeness
arose, questions that were evaluated in part by cross-referencing the SDSS
discoveries with previously-known magnetic white dwarfs in surveyed portions
of the sky.

Beginning with this edition, the strictly visual identification process was
augmented by an automated process originally developed to compile the SDSS DR3
white dwarf catalog (Eisenstein et al. 2005).  While that catalog was not
specifically designed to find magnetic white dwarfs, it recovered a reasonable
number of examples. Briefly, the automated procedure begins with a color cut in
$u-g$ and $g-r$ for point sources that have spectra not confidently classified
as extragalactic by the SDSS automated software. The spectra and photometry are
then fit to a temperature and surface gravity grid of (non-magnetic) pure
hydrogen and helium white dwarf atmosphere spectra using the {\tt autofit}
program (Kleinman et al. 2004).  Deviations from the best-fitting model are
computed for a variety of lines, including \ion{Ca}{2} $\lambda$3933,
\ion{He}{1} $\lambda$4471, \ion{He}{2} $\lambda$4686, H$\beta$, \ion{Mg}{1}b
$\lambda$5184, H$\alpha$, and the (1$-$0) C$_2$ Swan band near 5165~\AA. Stars
that present statistically significant deviations at these wavelengths or fail
one of a number of other quality assurance tests, such as an acceptable value
for the overall reduced $\chi^2$, are flagged for visual inspection.  Stars
with moderate to strong fields tend to trigger inspections either because of
poor overall fits or because of the above line tests.  Weak magnetic fields
(conservatively, those with $B\lesssim3$~MG, depending on object brightness)
may not perturb the spectra enough for the automated classifier to detect the
deviation; they remain among the non-magnetic DA and DB lists and require
visual recognition for discovery. We emphasize that in no case does the
automated process classify a star as magnetic; rather a spectrum is simply
indicated for future visual examination. Furthermore, the automated procedure
only contributes to the eventual catalog of magnetic stars; more than half of
the discoveries reported here originated from visual assessment of survey
spectra, an effort that is continuously under way.

\section{Results}

\subsection{Atmospheric Composition and Field Strength}

All but one of the 52 magnetic white dwarfs found in the SDSS database since
Paper I show hydrogen features, as listed in Table 1.  Fifty are
apparently strictly DAs, two are DQAs showing additional lines of atomic
carbon, and one shows molecular carbon features probably related to LHS 2229
and SDSS~J1333+0016 (Paper I)\footnote{Beyond these, three stars reported in
Paper I were rediscovered from new targetings of previously-observed SDSS
fields.  In Plate-MJD-Fiber designation, these are 710-52203-311 =
709-52205-120 = 411-51817-172 = SDSS~J0304$-$0025, 2049-53349-450 =
810-52326-392 = 415-51810-370 = SDSS~J0331+0045 = KUV~03292+0035, and the magnetic DB star 1898-53260-299 = 429-51820-311 = SDSS~J0142+1315.  Two stars
contained in Table 1 are rediscoveries of white dwarfs not previously
recognized as magnetic.}.  The full SDSS coordinate designation
($hhmmss.ss\pm ddmmss.s$, J2000) is provided in Table 1, though stars are
abbreviated by the first four digits of each coordinate in the following text
and figures. The table also contains the Plate-MJD-Fiber spectroscopic
identifier, International Atomic Time (TAI) date and time of the midpoint of
the spectroscopic observation, and comments regarding spectral appearance, data
release of the fiber spectrum, followup polarimetry, aliases, etc.  Polar field
strengths have been determined by visual comparison to model magnetic spectra
as described in Paper I.  To be consistent, a centered dipole has been assumed
for the field structure, even in cases where the fits are obviously not
well-represented by this morphology, and the inclination represents the viewing
angle relative to the dipole axis, with a typical uncertainty of 30$^\circ$.
Uncertainties in the inclination are larger for very faint or weak-line
objects, and inclinations are not quoted at all for fields $\gtrsim$500~MG,
where the often very smeared spectra reduce the number of points of detailed
comparison.

Derived polar field strengths for the best-fitting dipole range from 1.5~MG to
$\sim$1000~MG, and quoted values can generally be considered accurate to better
than 10\%.  The current list is distinguished by no less than 5 examples with
hydrogen features at $B_p>500$~MG:  SDSS~J0021+1502 (550~MG); SDSS~J1351+5419
(760~MG); SDSS~J1206+0813 ($\sim$830~MG); SDSS~J1003+0538 (900~MG), and
SDSS~J2346+3853 (1000~MG).  The latter star rivals PG~1031+234 (Schmidt et al.
1986) for the highest field strength yet detected on a white dwarf.
SDSS~J1351+5419 is a rediscovery of a previously-known magnetic star from the
Second Byurakan objective-prism QSO survey (Liebert et al. 1994).  Spectra of
the new high-field DAs comprise an instructive progression with field strength,
as shown in Figure 1.  For $B_p\gtrsim400$~MG, persistent lines appear at
$\sim$4200~\AA\ and 4550~\AA, formed at near-stationary points of the H$\beta$
transitions $2s0$ to $4f0$ and $2s0$ to $4f$$-$$1$, respectively.  Other
distinctive features at high fields are the pair of H$\alpha$ components ($2s0$
to $3p0$ and $2p$$-$$1$ to $3d$$-$$2$) that slowly cross in the region
$\sim$$6000-7000$~\AA, and the stationary $2p0$ to $3d$$-$$1$ component of
H$\alpha$ at $\sim$8500~\AA\ can be recognized in the brighter stars.

The existence of strong magnetic fields on SDSS~J2346+3853, SDSS~J0021+1502,
and SDSS~J1206+0813 was confirmed through circular (and in the former case
linear) spectropolarimetry.  At such strong fields, polarization is more a
property of the continuum than the lines (e.g., Schmidt et al. 1996), so the
polarization spectra are not highly structured, and the values entered in Table
1 are averages over the range $\lambda\lambda4200-8400$.  The modest
polarization of SDSS~J2346+3853, when compared with other very strongly
magnetic DA white dwarfs like PG~1031+234 (Schmidt et al. 1986), Grw
+70$^\circ$8247 (Angel et al. 1985), or SDSS~J2247+1456 (Paper I), suggests
that the field morphology on the new star may be unusually tangled.

Spectra of the two magnetic DQA stars are presented in Figure 2.
SDSS~J0236$-$0808 was observed as part of DR1 but was not recognized as
magnetic in Paper I.  The object is faint, the line splitting is only partially
resolved at $B_p=5$~MG, and the presence of carbon features probably confused
the visual identification.  The rather weak magnetic field is consistent with
an overall lack of circular polarization ($v=V/I<0.1\%$).  SDSS~J1328+5908 was
included among the 18 carbon-feature white dwarfs discovered in the SDSS
through 2003 (Liebert et al. 2003b).  Here we point out that the hydrogen lines
appear as clear Zeeman triplets in a modest magnetic field, as indicated in the
figure for $\langle B \rangle=8.5$~MG.  The spectra of \ion{C}{1} and
\ion{C}{2} have not yet been calculated in the Paschen-Back regime, but at this
field strength the transitions can be assumed to resemble normal triplets, with
any given ensemble possibly somewhat skewed and displaced to the blue by the
quadratic Zeeman effect.  The $\pi$ components of these features would be
expected to line up near zero-field wavelengths, and indeed there is a good
correspondence between several of the observed features and the principal lines
observed in atomic DQ white dwarfs (marked in Fig. 2, as taken from Liebert et
al. 2003b).

SDSS~J0954+0913 is an interesting example of a white dwarf dominated by
molecular absorption that may represent a ``transition'' object between normal
(C$_2$) DQ stars and the ``peculiar'' DQs discussed by Schmidt et al. (1995).
A comparison is made of an MMT spectrum of SDSS~J0954+0913 obtained with SPOL
on 2004 24 April 24 with spectra of SDSS~J2310$-$0057 (DQ; Harris et al. 2003)
and LHS~2229 (magnetic peculiar DQ; Schmidt et al. 1999) in Figure 3.  Both the
positions and shapes of the bands in SDSS~J0954+0913 are intermediate between
the other two stars, suggesting that the features in all three are the C$_2$
Swan bands.  If so, a C$_2$ identification (Bues 1999) for the displaced and
distorted bands in LP 790-29 (and SDSS~J1113+0146) might be called into
question, since these lack the ``scalloped'' appearance of the former stars.
The spectrum-added circular polarization for SDSS~J0954+0913, obtained from the
same MMT spectropolarimetric data, is $v=-0.93\%$.  This is significant but
less in magnitude than the $-$3.8\% found for LHS~2229.  The polarization
within the bands approaches +20\% in the latter object, but no such increase is
seen in SDSS~J0954+0913.  If, as has been suggested, LHS~2229 has a surface
magnetic field of $\sim$100~MG, it would appear that the value on
SDSS~J0954+0913 is substantially less.

Three additional objects whose survey spectra are displayed in Figure 4 deserve
mention here.  SDSS~J1007+1237 exhibits clear Zeeman patterns in its Balmer
series, but the structure in at least H$\alpha$ and H$\beta$ is more complex
than the model spectra predict for a polar field strength of $\sim$7 MG.  The
spectral energy distribution is also very odd, suggesting either a calibration
mismatch between the blue and red spectrograph channels or a broad depression
extending from $\sim$5800~\AA\ to at least H$\alpha$.  Spectropolarimetric
observations are planned of this faint target to investigate whether it might
be a magnetic white dwarf pair or a single star with very complex field
structure. The second star, SDSS~J1234+1248, also shows Zeeman triplets in a
field of $\sim$7~MG, but the $\pi$ components are both unusually sharp and
considerably deeper than the models predict; indeed deeper than any other
low-field DA in the sample.  Possible interpretations include a
magnetic/nonmagnetic DA pair (DAH+DA), or once again a distinctly nondipolar
field geometry.  In this case a large portion of the star would be covered by
very weak magnetic field ($\lesssim$1~MG).  An example of analogous field
structure at a somewhat weaker strength is WD~1953$-$011 (Maxted et al. 2000).
Finally, SDSS~J9044+5321 presents a challenge.  The Balmer lines are seen
at their rest wavelengths, however other features are clearly visible in the
spectrum.  This is strong evidence of a binary system.  Attempt to match the
additional lines with DQ or DB features have been unsuccessful.
A magnetic DA model with a polar field near 53~MG will cause the $\pi$ component
of H$\alpha$ to shift enough to match up with the blue wing on the observed
profile but the model is unable to reproduce the features seen in the H$\beta$
region.  Followup observations are needed to determine the true nature of
this system.

\subsection{Photometry and Stellar Parameters}

In Table 2 we collect the {\it psfmag\/} photometry of the new magnetic stars
from survey imaging data.  The final column contains a temperature estimate
derived from a comparison of a star's observed colors (primarily $u-g$ and
$g-r$) to those computed from spectral models for nonmagnetic DA white dwarfs
with various values of $\log g$ by Bergeron et al. (1995).  No temperature is
listed for SDSS J0954+0913 because the optical spectral energy distribution is
strongly affected by the deep molecular bands.  For SDSS J1508+3945, the
photometry led to inconsistent temperature estimates among the colors; indeed
the {\it psfmag\/} values do not track the fiber spectrum and differ by up to
0.6 mag from the ``fiber'' magnitudes.  In this case the quoted temperature is
based on the fiber magnitudes, which are provided in the table footnote.
Temperatures derived from broad-band colors are prone to errors because of a
magnetic field's effect on the continuum opacity and absorption lines, both of
which increase in importance with field strength.  Values whose uncertainties
are estimated to be larger than 1000~K, either because of a strong field or
because they lie off the locus in the model 2-color diagram, are flagged by a
colon in the table.  Derived temperatures span 7,000~K to $\sim$28,000~K, thus
occupying the same range as the bulk of nonmagnetic SDSS DA white dwarfs from
Kleinman et al. (2004), whose temperatures and gravities were derived by
line-profile fitting.  Unfortunately, the cool white dwarfs that comprise the
peak of the white dwarf luminosity function are not targeted by the survey
unless they have unusual spectral energy distributions.  This selection effect
likely contributes to the fact that nearly all of the magnetic discoveries from
the SDSS are of DA spectral type.

While the 2-color diagrams are most useful for temperature estimation, the
$u-g, g-r$ plot also offers some leverage on the stellar surface gravity
through the sensitivity of the $u-g$ color to the Balmer discontinuity (see,
e.g., Figure 1 of Harris et al. 2003).  Estimates for $\log g$ based on
broadband colors are not as accurate as those from the popular method of
fitting Balmer-series line profiles (Bergeron et al. 1992), but the
spectroscopic technique is rendered ineffective by the reduction in Stark
broadening in strong magnetic fields.  Even photometric estimates of $\log g$
for the SDSS magnetic examples are interesting as an ensemble because of
mounting evidence that the typical magnetic white dwarf has a higher gravity,
and is therefore more massive, than its nonmagnetic counterpart (Liebert et
al. 2003a and references therein).  The 52 stars from Table 2 that can be used
for this purpose bear out this effect, with the average gravity, $\av{\log g} =
8.31$, being higher than the value of 8.06 that resulted from an application of
the spectroscopic method to the 1888 nonmagnetic DA white dwarfs from DR1
(Kleinman et al. 2004).  
The comparison is complicated by the facts that the spectroscopic
technique seems to show a bias for high gravities below $T_{\rm eff}
= 10,000$~K (Kleinmann et al. 2004), and the photometric method is
subject to interstellar reddening.  Within the SDSS database, the DR2
and DR3 data are uncorrected for galactic extinction and we have not
applied any correction of our own in the course of our analysis.  By
comparison, the colors used in Kleinman et al. (2004) had full
Galactic reddening corrections applied.  The differences are
generally small, with a median $E(B-V) = 0.034$~mag (http://www.sdss.org/dr3/algorithms/spectrophotometry.html), but we note
that applying reddening corrections to the DR2/DR3 magnetic white
dwarf photometry would tend to {\it increase\/} the inferred
gravities for stars hotter than 10,000~K, and we would expect to see an 
even larger difference
between the average gravity of the magnetic and nonmagnetic samples.
Either of the above mean gravities are larger than the $\av{\log
g} = 7.91$ found by Bergeron et al.  (1992) for 129 hot DAs from the McCook \&
Sion (1987) catalog.  The difference corresponds to a typical mass excess for
the magnetic stars of $\sim$0.30~$M_\sun$, bringing the mean mass for the SDSS
magnetic white dwarfs near the $\sim$0.93~$M_\sun$ derived for the magnetic
samples of Liebert et al. (2003a).

In light of the above result, SDSS~J0933+1022 is a particularly glaring outlier
with a gravity estimate of only $\log g \sim 6.3$.  The estimate results from
an unusually large $u-g$ color of +0.57, but a low gravity is supported by the
lack of detectable wings to the Balmer lines in the survey spectrum shown in
Figure 4.  Unfortunately, magnetic effects on the profiles invalidates a
determination of the gravity by the spectroscopic method, and in any case that
technique exhibits an unexplained, systematic problem with fitting DA stars
cooler than about 11,000K, as discussed in Kleinman et al. (2004).  The 1.5~MG
polar field strength of SDSS~J0933+1022 is weak for a white dwarf, but
sufficient to rule out a main-sequence or horizontal-branch nature, while the
temperature of 8,500~K eliminates a subdwarf possibility and at the same time
accents the lack of an obvious spectral companion.  The star shares some
similarities with SDSS~J1234$-$0228, which was modeled by Liebert et al. (2004)
to have $\log g = 6.38 \pm 0.05$ and $T_{\rm eff} = 17,500$~K, and interpreted
as a helium-composition white dwarf with $M = 0.18-0.19$~$M_\sun$ resulting
from some type of evolution with a yet undiscovered binary companion.  However,
considering both its surface temperature and gravity, SDSS~J0933+1022 perhaps
best resembles the companion to the pulsar J1012+5307, estimated by Van
Kerkwijk et al. (1996) and Callanan et al.  (1998) to have $T_{\rm eff} \sim
8,600$~K and $M = 0.16$~$M_\sun$.  The star demands followup study, both to
confirm its low gravity and to look for a massive cool white dwarf, very
low-mass main-sequence, or dead pulsar companion.

\subsection{Issues of Completeness}

Including the additions from this paper, a total of 40 magnetic white dwarfs
were identified in the first 1360 deg$^2$ of the SDSS released as DR1.  This
should be compared with only 49 new magnetic stars discovered in the additional
2828 deg$^2$ covered by DR2 and DR3 (22 discoveries were reported in Paper I).
There is actually good reason to believe that our current identification
procedures are more effective than the strictly visual method of Paper I, once
a spectrum is obtained.  The machine selection algorithm is not only far more
efficient but also more uniform, and the 3 DR1 stars overlooked in the course
of preparing Paper I but recovered here attest to its utility. Of course, the
areal density of stars varies with location on the sky, but the considerable
discrepancy in discovery rate of magnetic stars suggests that improvements in
the photometric reduction pipeline and continued refinement of the QSO
targeting criteria have reduced the chance that an interloper is selected for a
spectroscopic fiber.  Comparison of the locations of magnetic white dwarfs in
the SDSS color planes for the DR1 and DR2/DR3 samples reveal no regions
populated solely by DR1 objects, but the photometric scatter around the locus
is found to be slightly reduced in the later sample for the shorter wavelength
filter bands.  More direct evidence for a reduction in the number of stellar
targetings because of improvements in the QSO targeting criteria is provided by
the fact that, of the 8 previously-known magnetic white dwarfs that lie in
regions of the sky covered by DR2/DR3, only two - SDSS~J1214$-$0234 = LHS 2534
(reported in Paper I) and SDSS~J1351+5419 = SBS1349+5434 - were recovered in
the survey.  Admittedly, G99-37, HS1412+6115, and GD 185 exceed the brightness
limit for a spectroscopic fiber, but the three magnetic DAs LHS 2273 (18~MG),
PG 1312+098 (10~MG), and G62-46 (7.4~MG) are all within the limits of the SDSS.
None was targeted for spectroscopy. This result should be contrasted with the
fact that 5 of the 6 qualifying stars in DR1 were recovered and reported in
Paper I.

The implied incompleteness of recent (and forthcoming) editions of the survey
raises concerns over the effects that a differential selection bias might have
on the derived distributions of magnetic stars as functions of field strength
or temperature, and thus on inferences that may be drawn from these
distributions.  However, the breakdown of new identifications from this paper,
per decade of field strength, is:  1$-$10 MG, 13; 10$-$100 MG, 11; 100$-$1000
MG, 3.  Within the sampling statistics, these are consistent with the
corresponding numbers 23:28:9 for stars from Paper I, and with the distribution
shown in Figure 8 of that paper for all 116 magnetic white dwarfs known at the
time.  The question of completeness for magnetic white dwarfs discovered from
the SDSS will be revisited when the survey is concluded.

\acknowledgements{Funding for the creation and distribution of the SDSS Archive
has been provided by the Alfred P. Sloan Foundation, the Participating
Institutions, the National Aeronautics and Space Administration, the National
Science Foundation, the U.S. Department of Energy, the Japanese Monbukagakusho,
and the Max Planck Society. The SDSS Web site is http://www.sdss.org/.  The
SDSS is managed by the Astrophysical Research Consortium (ARC) for the
Participating Institutions. The Participating Institutions are The University
of Chicago, Fermilab, the Institute for Advanced Study, the Japan Participation
Group, The Johns Hopkins University, the Korean Scientist Group, Los Alamos
National Laboratory, the Max-Planck-Institute for Astronomy (MPIA), the
Max-Planck-Institute for Astrophysics (MPA), New Mexico State University,
University of Pittsburgh, University of Portsmouth, Princeton University, the
United States Naval Observatory, and the University of Washington.  Support for
the study of magnetic stars and stellar systems at Steward Observatory is
provided by NSF grant AST 03-06080.}

\begin{deluxetable}{lrlrrrl}
\tabletypesize{\scriptsize}

\rotate


\tablecaption{NEW SDSS MAGNETIC WHITE DWARFS}

\tablenum{1}
\tablewidth{9.in}
\setlength{\tabcolsep}{0.06in}

\tablehead{\colhead{Star} & \colhead{Plate-MJD-Fiber} & \colhead{TAI Date} & \colhead{TAI} & \colhead{$B_p$} & \colhead{$i$} & \colhead{Comments} \\
\colhead{(SDSS+)} & \colhead{} & \colhead{} & \colhead{} & \colhead{(MG)} & \colhead{(deg)} & \colhead{} }

\startdata
J002129.00+150223.7 & 753-52233-432 & 2001 Nov 20 & 4:27 & 550 & \ldots & $v=+0.91\%$    \\
J021148.22+211548.2 & 2046-53327-048 & 2004 Nov 11 & 6:13 & 210 & 90 & post-DR3 \\
J023609.40$-$080823.9 & 455-51909-474 & 2000 Dec 31 & 3:28 & 5 & 90 & DQA; $v=+0.01\%$; in DR1 \\
J074850.48+301944.8 & 889-52663-507 & 2003 Jan 24 & 4:54 & 10 & 60 &  \\
J080440.35+182731.0 & 2081-53357-442 & 2004 Dec 18 & 7:19 & 49 & 30 & post-DR3\\
J080502.29+215320.5 & 1584-52943-132 & 2003 Oct 30 & 11:17 & 5 & 60 & post-DR3 \\
J080938.10+373053.8 & 758-52253-044 & 2001 Dec 7 & 7:55 & 40 & 30 &  \\
J081648.71+041223.5 & 1184-52641-329 & 2003 Jan 2 & 8:23 & 10: & 30: &  \\
J082835.82+293448.7 & 1207-52672-635 & 2003 Feb 2 & 4:30 & 30 & 90: &  \\
J084008.50+271242.7 & 1587-52964-059 & 2003 Nov 21 & 9:35 & 10 & 60 & post-DR3 \\
J090632.66+080716.0 & 1300-52973-148 & 2003 Nov 30 & 11:31 & 10 & 90 & post-DR3\\
J090746.84+353821.5 & 1212-52703-187 & 2003 Mar 5 & 4:28 & 15 & 60 &  \\
J091124.68+420255.9 & 1200-52668-538 & 2003 Jan 23 & 10:05 & 45 & 60 &  \\
J091437.40+054453.3 & 1193-52652-481 & 2003 Jan 11 & 9:19 & 9.5 & 90 &  \\
J093356.40+102215.7 & 1303-53050-525 & 2004 Feb 12 & 9:17 & 1.5 & 60: &  = SDSS 1303-53047-525; post-DR3\\
J093447.90+503312.2 & 901-52641-373 & 2003 Jan 2 & 9:38 & 9.5 & 60 &  \\
J094458.92+453901.2 & 1202-52672-577 & 2003 Feb 2 & 8:24 & 14 & 90 & \\
J095442.91+091354.4 & 1306-52996-004 & 2004 Dec 23 & 10:54 & \ldots & \ldots & mol. DQ; $v=-0.93\%$; post-DR3 \\
J100356.32+053825.6 & 996-52641-295 & 2003 Jan 2 & 10:50 & 900: & \ldots &\\
J100715.55+123709.5 & 1745-53061-313 & 2004 Feb 26 & 4:05 & 7 & 60 & post-DR3\\
J101529.62+090703.8 & 1237-52762-533 & 2003 May 3 & 4:50 & 5: & 90: &\\
J111812.67+095241.4 & 1222-52763-477 & 2003 May 4 & 4:37 & 6 & 60: &\\
J112924.74+493931.9 & 966-52642-474 & 2003 Jan 3 & 10:12 & 5 & 60 &\\
J113756.50+574022.4 & 1311-52765-421 & 2003 May 6 & 5:32 & 9 & 60: &\\
J114829.00+482731.2 & 1446-53080-324 & 2004 Mar 16 & 6:00 & 33 & 90 & post-DR3\\
J120150.10+614257.0 & 778-52337-264 & 2002 Mar 4 & 8:22 & 20 & 90 &\\
J120609.80+081323.7 & 1623-53089-573 & 2004 Mar 25 & 6:24 & 830: & \ldots & $v=+0.41\%$; post-DR3 \\
J120728.96+440731.6 & 1369-53089-048 & 2004 Mar 25 & 7:53 & 2.5 & 90 & post-DR3\\
J122249.14+481133.1 & 1451-53117-582 & 2004 Apr 22 & 5:56 & 8 & 90: & post-DR3\\
J122401.48+415551.9 & 1452-53112-181 & 2004 Apr 17 & 6:30 & 23:  & 60 & post-DR3\\
J123414.11+124829.6 & 1616-53169-423 & 2004 Jun 13 & 3:57 & 7 & 60: & WD1231+130=LBQ 1231+1305; post-DR3\\
J124806.38+410427.2 & 1456-53115-190 & 2004 Apr 20 & 4:54 & 8 & 90 & post-DR3\\
J125044.42+154957.4 & 1770-53171-530 & 2004 Jun 15 & 4:16 & 20 & 60 & post-DR3\\
J125416.01+561204.7 & 1318-52781-299 & 2003 May 22 & 5:54 & 52 & 30 &\\
J132002.48+131901.6 & 1773-53090-011 & 2004 Mar 26 & 10:23 & 5 & 60 & post-DR3\\
J132858.20+590851.0 & 959-52411-504 & 2002 May 17 & 5:56 & 18 & 90 & DQA\\
J135141.13+541947.4 & 1323-52797-293 & 2003 Jun 2 & 7:25 & 760 & 20 & SBS1349+5434 \\
J142703.40+372110.5 & 1381-53089-182 & 2004 Mar 25 & 11:12 & 30 & 60: & post-DR3\\
J143218.26+430126.7 & 1396-53112-338 & 2004 Apr 17 & 10:17 & 1.5 & 90 & post-DR3\\
J143235.46+454852.5 & 1288-52731-449 & 2003 Apr 2 & 7:01 & 10 & 30 &\\
J145415.01+432149.5 & 1290-52734-469 & 2003 Apr 5 & 10:26 & 5 & \ldots &\\
J150813.20+394504.9 & 1398-53146-633 & 2004 May 21 & 8:01 & 20 & 90 & WD1506+399 = CBS 229; post-DR3\\
J151130.20+422023.0 & 1291-52735-612 & 2003 Apr 6 & 10:12 & 12 & 60 &\\
J164357.02+240201.3 & 1414-53121-191 & 2004 Apr 23 & 11:16 & 4 & 90 & post-DR3\\
J164703.24+370910.3 & 818-52395-026 & 2002 May 1  & 9:38 & 2: & 90: &\\
J165029.91+341125.5 & 1175-52791-482 & 2003 Jun 1 & 9:53 & 3: & 0: &\\
J170400.01+321328.7 & 976-52413-319 & 2002 May 17 & 10:55 & 5 & 90: &\\
J171556.29+600643.9 & 354-51792-318 & 2000 Sep 5 & 4:27 & 4.5 & 60 & in DR1\\
J214900.87+004842.8 & 1107-52968-374 & 2003 Nov 22 & 1:50 & 10 & 60 & post-DR3\\
J231951.73+010909.3 & 382-51816-565 & 2000 Sep 29 & 4:44 & 1.5: & 90: & in DR1\\
J234605.44+385337.7 & 1883-53265-272 & 2004 Sep 17 & 9:22 & 1000 & \ldots & $v=-1.51\%$; $P=1.33\%, \theta=164^\circ$; post-DR3\\
J234623.69$-$102357.0 & 648-52559-142 & 2002 Oct 12 & 3:56 & 2.5 & 90: & \\
\enddata




\end{deluxetable}

\clearpage

\begin{deluxetable}{lcccccr}




\tablecaption{PHOTOMETRY OF SDSS MAGNETIC WHITE DWARFS}

\tablenum{2}
\tablewidth{5.25in}
\setlength{\tabcolsep}{0.1in}

\tablehead{\colhead{Star} & \colhead{$u$} & \colhead{$g$} & \colhead{$r$} & \colhead{$i$} & \colhead{$z$} & \colhead{$T_{eff}$} \\
\colhead{(SDSS+)} &  &  &  &  &  & \colhead{(K)} }

\startdata
J002129.00+150223.7  & 18.69 & 18.18 & 17.97 & 17.93 & 17.99 & 7000 \\
J021148.22+211548.2 & 17.26 & 17.07 & 17.21 & 17.41 & 17.63 & 12000 \\
J023609.40$-$080823.9  & 19.96 & 19.76 & 19.78 & 19.84 & 19.95 & 10000 \\
J074850.48+301944.8 & 17.39 & 17.58 & 17.87 & 18.49 & 18.56 & 22000: \\
J080440.35+182731.0 & 18.45 & 18.13 & 18.25 & 18.32 & 18.55 & 11000  \\
J080502.29+215320.5 & 18.25 & 18.60 & 18.98 & 19.37 & 19.73 & 28000: \\
J080938.10+373053.8 & 19.32 & 19.02 & 19.26 & 19.49 & 19.76 & 14000 \\
J081648.71+041223.5 & 20.74 & 20.40 & 20.52 & 20.63 & 21.11 & 11500 \\
J082835.82+293448.7 & 19.67 & 19.74 & 20.05 & 20.32 & 20.31 & 19500 \\
J084008.50+271242.7 & 19.53 & 19.18 & 19.35 & 19.54 & 19.87 & 12250 \\
J090632.66+080716.0 & 18.62 & 18.65 & 18.82 & 19.11 & 19.32 & 17000: \\
J090746.84+353821.5 & 19.76 & 19.63 & 19.92 & 20.27 & 20.48 & 16500 \\
J091124.68+420255.9 & 19.17 & 18.85 & 18.91 & 19.02 & 19.24 & 10250 \\
J091437.40+054453.3 & 17.44 & 17.33 & 17.64 & 17.89 & 18.16 & 17000 \\
J093356.40+102215.7 & 19.34 & 18.77 & 18.73 & 18.71 & 18.73 & 8500 \\
J093447.90+503312.2 & 19.19 & 18.81 & 18.80 & 18.85 & 19.07 & 8900 \\
J094458.92+453901.2 & 20.01 & 19.95 & 20.14 & 20.35 & 20.74 & 15500: \\
J095442.91+091352.4 & 20.07 & 20.16 & 19.71 & 19.67 & 19.97 & \ldots \\
J100356.32+053825.6 & 17.96 & 18.12 & 18.48 & 18.65 & 18.90 & 23000 \\
J100715.55+123709.5 & 18.78 & 18.77 & 19.05 & 19.32 & 19.57 & 18000 \\
J101529.62+090703.8 & 19.10 & 18.59 & 18.41 & 18.39 & 18.49 & 7200 \\
J111812.67+095241.4 & 19.17 & 18.75 & 18.85 & 19.04 & 19.19 & 10500 \\
J112924.74+493931.9 & 18.42 & 18.00 & 18.08 & 18.15 & 18.34 & 10000 \\
J113756.50+574022.4 & 17.32 & 16.85 & 16.75 & 16.75 & 16.77 & 7800 \\
J114829.00+482731.2 & 17.88 & 18.16 & 18.58 & 18.84 & 19.18 & 27500 \\
J120150.10+614257.0 & 18.80 & 18.48 & 18.57 & 18.60 & 18.78 & 10500 \\
J120609.80+081323.7 & 19.27 & 19.04 & 19.22 & 19.42 & 19.58 & 13000 \\
J120728.96+440731.6 & 19.44 & 19.33 & 19.62 & 19.97 & 20.52 & 16750 \\
J122249.14+481133.1 & 19.11 & 18.72 & 18.72 & 18.79 & 18.97 & 9000 \\
J122401.48+415551.9 & 19.27 & 18.96 & 18.98 & 19.06 & 19.25 & 9500 \\
J123414.11+124829.6 & 17.85 & 17.38 & 17.32 & 17.36 & 17.49 & 8200 \\
J124806.38+410427.2 & 18.40 & 17.93 & 17.71 & 17.72 & 17.79 & 7000 \\
J125044.42+154957.4 & 18.64 & 18.26 & 18.32 & 18.38 & 18.20 & 10000 \\
J125416.01+561204.7 & 19.36 & 18.98 & 19.20 & 19.33 & 19.59 & 13250 \\
J132002.48+131901.6 & 19.84 & 19.49 & 19.82 & 20.07 & 20.19 & 14750 \\
J132858.20+590851.0 & 17.63 & 17.85 & 18.25 & 18.55 & 18.89 & 25000 \\
J135141.13+541947.4 & 16.66 & 16.40 & 16.55 & 16.72 & 16.94 & 12000 \\
J142703.40+372110.5 & 17.55 & 17.57 & 17.91 & 18.18 & 18.43 & 19000 \\
J143218.26+430126.7 & 18.86 & 18.99 & 19.41 & 19.71 & 20.15 & 24000 \\
J143235.46+454852.5 & 20.06 & 19.94 & 20.24 & 20.49 & 20.52 & 16750 \\
J145415.01+432149.5 & 19.48 & 19.06 & 19.21 & 19.34 & 19.72 & 11500 \\
J150813.20+394504.9 & 17.35 & 17.90 & 17.75 & 18.03 & 18.51 & 17000\tablenotemark{a} \\
J151130.20+422023.0 & 18.20 & 17.98 & 18.01 & 18.19 & 18.46 & 9750 \\
J164357.02+240201.3 & 19.43 & 19.27 & 19.58 & 19.82 & 20.40 & 16500 \\
J164703.24+370910.3 & 17.80 & 17.64 & 17.92 & 18.21 & 18.50 & 16250 \\
J165029.91+341125.5 & 19.13 & 18.78 & 18.83 & 18.98 & 19.09 & 9750 \\
J170400.01+321328.7 & 20.30 & 20.43 & 20.83 & 21.07 & 21.91 & 23000 \\
J171556.29+600643.9 & 19.70 & 19.35 & 19.59 & 19.73 & 19.78 & 13500 \\
J214900.87+004842.8 & 20.12 & 19.87 & 19.96 & 20.19 & 20.05 & 11000 \\
J231951.73+010909.3 & 18.82 & 18.50 & 18.44 & 18.52 & 18.61 & 8300 \\
J234605.44+385337.7 & 18.64 & 18.89 & 19.28 & 19.48 & 19.85 & 26000 \\
J234623.69$-$102357.0 & 18.77 & 18.45 & 18.40 & 18.44 & 18.59 & 8500 \\
\enddata


\tablenotetext{a}{Temperature based on $ugriz$ ``fiber'' magnitudes of 17.40,
17.30, 17.60, 17.88, 18.21, respectively.}



\end{deluxetable}

\clearpage

\begin{figure}
\includegraphics{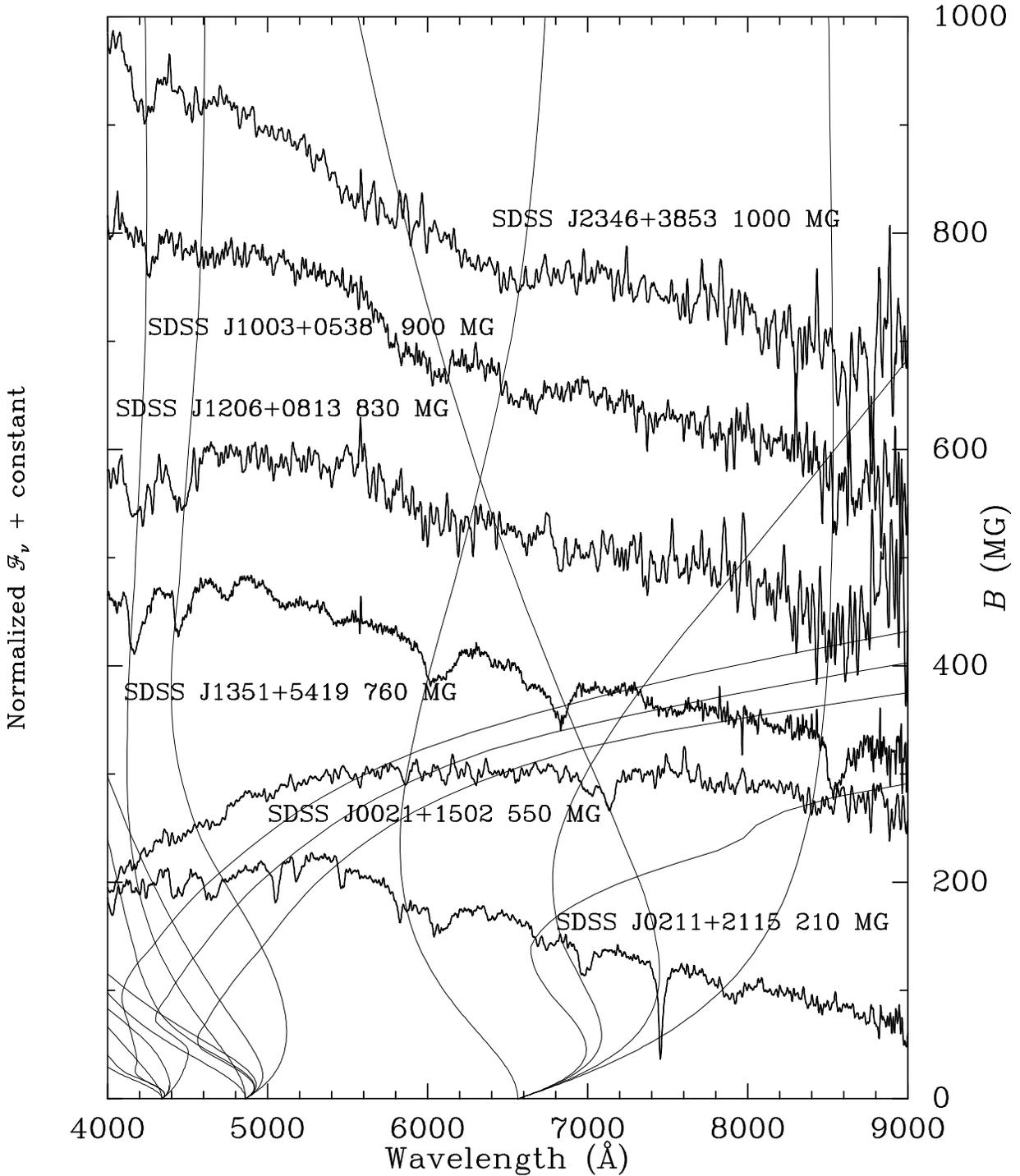}
\vspace{7.25truein}

\figcaption{Progression of new strongly-magnetic DA white dwarfs showing
features of hydrogen in fields up to 1000~MG, and compared with computed
wavelengths for transitions that become stationary or execute turnarounds in
the field strength range of interest.  Persistent at the highest fields is the
pair of H$\beta$ components that become nearly stationary in the range
4200$-$4600~\AA\ and the two H$\alpha$ components that wander between
6000$-$7000~\AA. Quoted field strengths are estimated polar values for the
best-fit assumed dipole model.  Spectra have been smoothed to a resolution of
$\sim$8~\AA, scaled, and shifted for display purposes.}
\end{figure}

\clearpage

\begin{figure}
\includegraphics{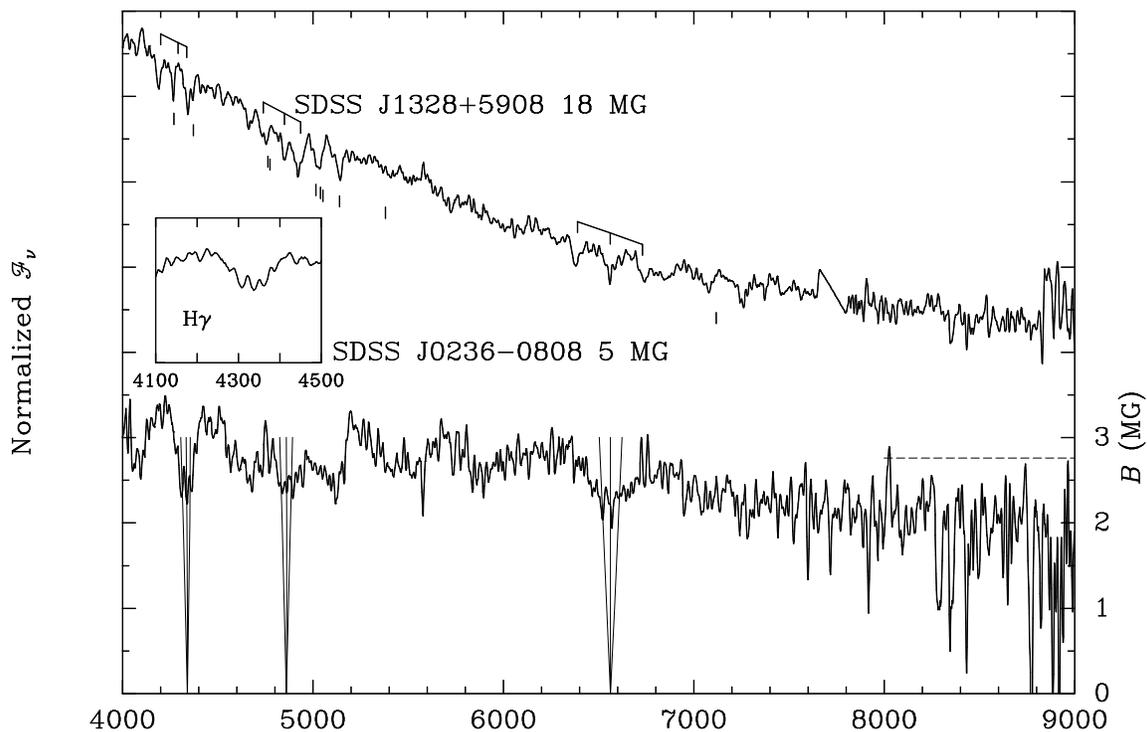}
\vspace{1.5truein}

\figcaption{Two unusual magnetic DQA (hydrogen + carbon) white dwarfs.  For
SDSS~J0236$-$0808, Zeeman triplets in the Balmer lines are legible by
comparison with the theoretical curves shown, with field strength indicated
along the right ordinate.  The largely blended molecular and atomic carbon
features render a somewhat choppy appearance to the spectrum.  SDSS~J1328+5908
is at a field strength where all features should resemble Zeeman triplets, but
carbon has not been calculated in this regime, so we simply point out that the
$\pi$ components of split features should be found near the zero-field
wavelengths of the same lines (indicated by tick marks below the observed
spectrum). The inset displays the \hgamma\ region for SDSS~J0236$-$0808 at a
larger scale, and a horizontal dashed line indicates the zero-flux level for
SDSS~J1328+5908.
}
\end{figure}

\clearpage

\begin{figure}
\includegraphics{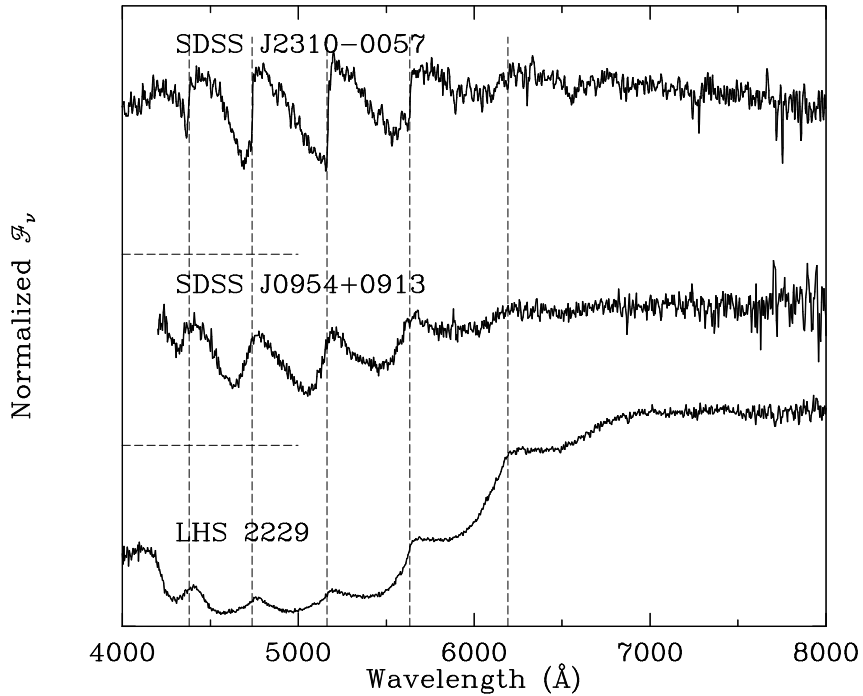}
\vspace{1.0truein}

\figcaption{Spectral comparison between SDSS~J0954+0913, the normal (C$_2$
Swan band) molecular white dwarf SDSS~J2310$-$0057, and the peculiar magnetic
DQ star LHS 2229. At $v=-0.93\%$, circular polarization of SDSS~J0954+0913 has
been measured to be significant, but it is smaller in magnitude than the
$v=4-20\%$ seen in LHS~2229, supporting an interpretation for SDSS~H0954+0913
as a intermediate object with a modest magnetic field ($B<100$~MG). Horizontal dashed lines indicate the zero-flux levels for successive spectra.}
\end{figure}

\clearpage

\begin{figure}
\vspace{2in}
\includegraphics{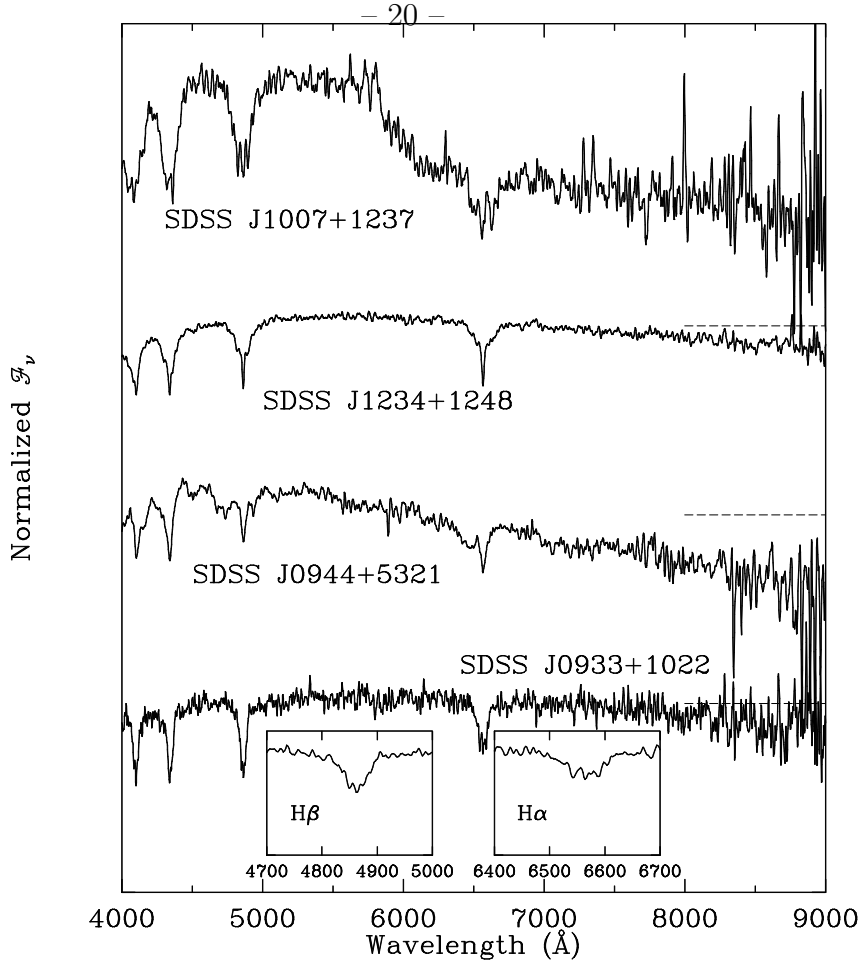}
\vspace{1.5truein}

\figcaption{Survey spectra of four unusual examples.  SDSS~J1007+1237 and
SDSS~J1234+1248 both have polar fields $B_p\sim$7~MG, but the former shows a
distorted spectral energy distribution, while the latter exhibits conspicuously
deep $\pi$ components suggestive of a nonmagnetic companion or a strongly
nondipolar field structure over the surface of the star.
We note that the distortion in the spectrum of SDSS~J1007+1237 may be due
to a calibration mismatch between the blue and red spectrograph channels.
The odd H$\alpha$ profile  of SDSS~J0944+5321 is suggestive of a companion
but DQ, DB, and magnetic DA models have all been unsuccessful in reproducing
the features seen in the H$\beta$ region.
Photometry of SDSS~J0933+1022 implies a very low gravity, $\log g < 7$,
suggesting a history involving a close binary companion.  Insets display the
\hbeta\ and \halpha\ regions for SDSS~J0933+1022 at a larger scale, and
horizontal dashed lines indicate the zero-flux levels for successive spectra.
}
\end{figure}

\end{document}